\documentstyle[epsfig]{aipproc}

\begin{document}
\title{Modeling the April 1997 flare\\ of Mkn~501}

\author{A. M\"ucke and R.J. Protheroe}
\address{The University of Adelaide\\
Dept. of Physics \& Mathematical Physics\\
Adelaide, SA 5005, Australia}

\maketitle

\begin{abstract}
The April 1997 giant flare of Mkn~501 is modelled by a stationary 
Synchrotron-Proton-Blazar model. 
Our derived model parameters are consistent with X-ray-to-TeV-data
in the flare state and diffusive shock acceleration of $e^-$ and $p$ in a Kolmogorov/Kraichnan
turbulence sprectrum. While the emerging pair-synchrotron cascade spectra initiated by photons from $\pi^0$-decay and
electrons from $\pi^\pm\rightarrow\mu^\pm\rightarrow e^\pm$-decay turn out to be relatively featureless,
$\mu^\pm$ and
$p$ synchrotron radiation and their cascade radiation
produce a double-humped spectral energy distribution. 
For the present model we
find $p$ synchrotron radiation to dominate the TeV emission, while 
the contribution from the synchrotron radiation
of the pairs, produced by the high energy hump, is only minor.

\end{abstract}


\section{The co-acceleration scenario}

With its giant outburst in 1997,
emitting photons up to $24$~TeV and $0.5$~MeV in the $\gamma$-ray and X-ray bands,
Mkn~501 has proved to be the most extreme TeV-blazar observed so far
(e.g. Catanese et al 1997, Pian et al 1997, Aharonian et al 1999).

In this paper, we consider the April 1997 flare of Mkn~501 in the light of a modified 
version of the Synchrotron Proton Blazar model (SPB) (Mannheim 1993), and present a preliminary model fit.

In the model, shock accelerated protons ($p$) interact
in the synchrotron photon field generated by the electrons ($e^-$) co-accelerated
at the same shock.
This scenario may put constraints on the maximum achievable
particle energies.

The usual process considered for accelerating charged particles in the plasma jet is
diffusive shock acceleration (see e.g. Drury 1983, Biermann \& Strittmatter 1987).
If the particle spectra are cut off due to synchrotron losses, the ratio of the  
maximum particle energies $\gamma_{p,max}/\gamma_{e,max}$
can be derived
by equating $t_{acc,p}/t_{acc,e} = t_{syn,p}/t_{syn,e}$, with
$t_{syn,p}$ and $t_{syn,e}$ being the synchrotron loss time scales for $p$ and $e^-$, respectively. We find that for shocks of compression ratio 4 
(see M\"ucke \& Protheroe 1999 for a detailed derivation)
\begin{equation}
\frac{\gamma_{p,\rm{max}}}{\gamma_{e,\rm{max}}} \leq \frac{m_p}{m_e} (\frac{m_p}{m_e})^{\frac{2(\delta-1)}{3-\delta}} \sqrt{\frac{F(\theta,\eta_{e,\rm{max}})}{F(\theta,\eta_{p,\rm{max}})}} =  
\frac{m_p}{m_e} \sqrt{\frac{\eta_{e,\rm{max}} F(\theta,\eta_{e,\rm{max}})}{\eta_{p,\rm{max}} F(\theta,\eta_{p,\rm{max}})}}
\end{equation}
where the ``=''-sign corresponds to
synchrotron loss, and the ``$<$''-sign to adiabatic loss determining the maximum energies. 
$\delta$ is the power law index of the magnetic turbulence spectrum ($\delta = 5/3$: Kolmogorov turbulence, $\delta = 3/2$: Kraichnan turbulence, and $\delta=1$ corresponds
to Bohm diffusion). $\eta_{e,\rm{max}}$ is the mean free path at maximum energy in units
of the particle's gyroradius and $F(\theta,\eta_{e,\rm{max}})$ takes account
of the shock angle $\theta$ (Jokipii 1987).
The ratio
$F(\theta,\eta_{e,\rm{max}}) \eta_{e,\rm{max}}/F(\theta,\eta_{p,\rm{max}}) \eta_{p,max}$
can be constrained by the variability time scale $t_{\rm{var}}$, requiring 
$t_{var} D \geq t_{acc,p,max}$ ($D$ = Doppler factor, $t_{acc,p,max}$ = acceleration time scale
at maximum particle energy) for a given parameter combination.
As an example, we adopt $D=10$, $B=20$G and $t_{\rm{var}}=2$~days.
Eq.~1 then restricts for these parameters the ratio of the allowed maximum
particle energies to the range below the solid lines shown in Fig.~1. Points exactly on this line
correspond to synchrotron-loss limited particle spectra which are
accelerated with exactly the variability time scale.

\vspace*{-.5cm}

\begin{figure}[h] 
\begin{minipage}[t]{3.55in}
\centerline{\epsfig{file=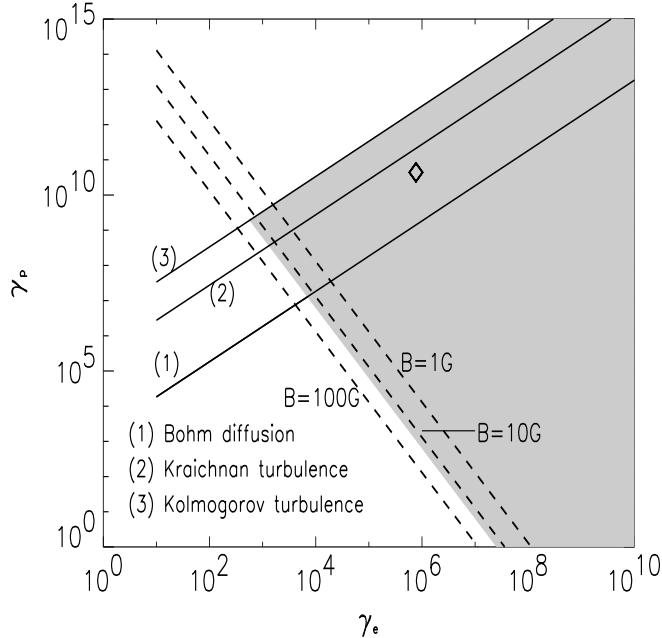,height=3.5in,width=3.5in}}
\end{minipage}
\hfill
\begin{minipage}[t]{2.2in}
\vspace*{-3.3in}
\caption{Allowed parameter space (shaded area) for $\gamma_{p,\rm{max}}$, $\gamma_{e,\rm{max}}$
in the SBP-model for typical TeV-blazar parameters
(B=20~G, D=10, $u_1=0.5c$, $\beta=1$, $t_{\rm{var}}=2$days) and for different magnetic turbulence spetra $I(k) \propto k^{-\delta}$. The diamond symbol corresponds
to the Mkn~501-model presented below.}
\end{minipage}
\label{fig1}
\end{figure}

\vspace*{-1cm}

In hadronic models $\pi$ photoproduction is essential for $\gamma$-ray production. The threshold
of this process is given by $\epsilon_{\rm{max}} \gamma_{p,\rm{max}} = 0.0745$~GeV where
$\epsilon_{\rm{max}}$ is the maximum photon energy of the synchrotron target field.  
Inserting $\epsilon_{\rm{max}} = 3/8 \gamma_{e,\rm{max}}^2 B/(4.414\times 10^{13}\rm{G})~m_e c^2$ into the threshold condition, we find

\vspace*{-.3cm}

$$
\gamma_{p,\rm{max}} \geq 1.72 \cdot 10^{16} (\frac{B}{\rm{Gauss}})^{-1} \gamma_{e,\rm{max}}^{-2} 
$$
which is shown in Fig.~1 as dashed line for various magnetic field strengths.
Together with Eq.~1, the allowed range of maximum particle energies is then restricted
to the shaded area in Fig.~1.


\section{The Mkn~501 flare in the Synchrotron Proton Blazar (SPB) Model}

We assume the parameters used in Fig.~1, and that the co-accelerated $e^-$ produce the {\underline{observed}} synchrotron spectrum, 
unlike in previous SPB models, and this is
the target radiation field for the $p\gamma$-interactions.
This synchrotron spectrum, and its hardening with rising flux, 
has recently been convincingly reproduced by a shock
model with escape and synchrotron losses (Kirk et al 1998).
We use 
the Monte-Carlo technique for particle production/cascade development, which allows us to use exact 
cross sections.


\begin{figure}[h] 
\begin{minipage}[t]{3.55in}
\centerline{\epsfig{file=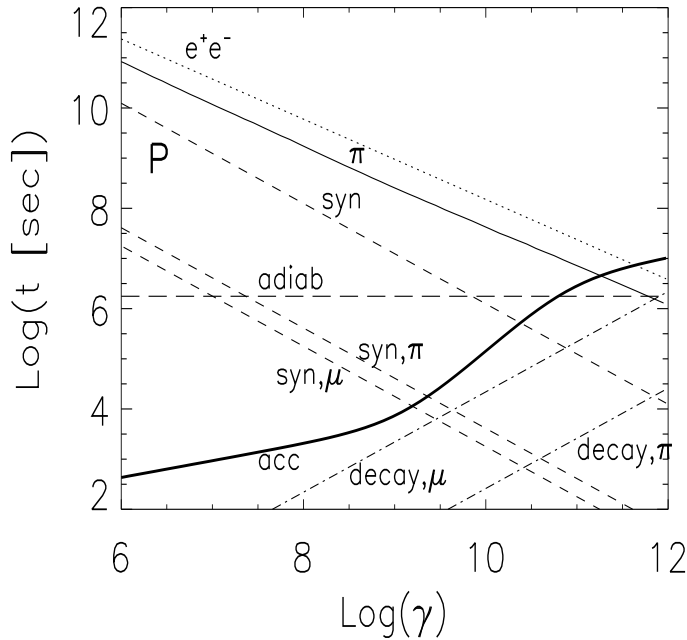,height=3.5in,width=3.7in}}
\end{minipage}
\hfill
\begin{minipage}[t]{2.25in}
\vspace*{-3.6in}
\caption{
Mean energy loss time of $p$ for synchrotron radiation (syn), $\pi$-photoproduction
($\pi$) and Bethe-Heitler~pair~production ($e^\pm$), 
and for $\pi^\pm$- and $\mu^\pm$-synchrotron radiation (syn $\pi$, syn $\mu$)
for $B=19.6$~G
with their mean decay time scales (decay $\pi$, decay $\mu$) in the
SPB-model. The acceleration time scale (acc), based on Kolmogorov turbulence, is calculated for 
$u_1 = 0.5c$, $\eta_p = 40$ and shock angle $\theta = 85^\circ$. Its curvature reflects the influence of
the shock angle. The adiabatic loss time (adiab) is assumed to be 
$R/u_1 \approx D t_{var}$.
All quantities are in the jet frame.}
\end{minipage}
\label{fig2}
\end{figure}

\vspace*{-1cm}

For simplicity we represent the observed synchrotron spectrum (target photon field for the 
$p\gamma$-collisions) as a broken power law in the jet frame with photon power law index 1.4
below the break energy of 0.2~keV, and index 1.8 up to 50~keV.

The variability time scale restricts the radius $R$ of the emission region.
For our model we use 
$t_{\rm{var,x}} \approx 2$~days (Catanese et al 1997), and find $R\approx 2.6\times 10^{16}$cm
for $D=10$, $B=19.6$~G. 
With these parameters the $\gamma\gamma$-pair
production optical depth reaches unity for $\approx 25$~TeV photons.

Our model considers photomeson production (simulated using
SOPHIA, M\"ucke et al 1999), Bethe-Heitler pair production (simulated using the code of 
Protheroe \& Johnson 1996),
$p$ synchrotron radiation and adiabatic losses due to jet expansion.
The mean energy loss and acceleration time scales 
are presented in Fig.~2. 

Synchrotron losses, which turn out to be at least as important as losses due to 
$\pi$ photoproduction for the assumed 2~day variability, limit the injected $p$ spectrum
$\propto \gamma_p^{-2}$ to
$2\leq\gamma_p\leq 4.4\times 10^{10}$. This leads to 
a $p$ energy density
$u_p \approx 0.2~\rm{TeV/cm}^3$, which is bracketed by the photon energy density 
$u_{\rm{target}} \approx 0.01~\rm{TeV/cm}^3$, and a magnetic field energy density 
$u_B \approx 9.5~\rm{TeV/cm}^3$. With $u_B \gg u_{\rm{target}}$ significant Inverse Compton radiation
from the co-accelerated $e^-$ is not expected.

Rachen \& Meszaros (1998) noted the
importance of synchrotron losses of $\mu^\pm$- (and $\pi^\pm$-) prior to their decay in AGN jets
and GRBs. 
For the
present model, the critical Lorentz factors $\gamma_{\mu} \approx 3 \times 10^9$ and $\gamma_{\pi} \approx 4 \times 10^{10}$,
above which synchrotron losses dominate above decay, lie well below the maximum particle energy for $\mu^\pm$,
while $\pi^\pm$-synchrotron losses can be neglected due to the shorter decay time.


\begin{figure}[h] 
\epsfig{file=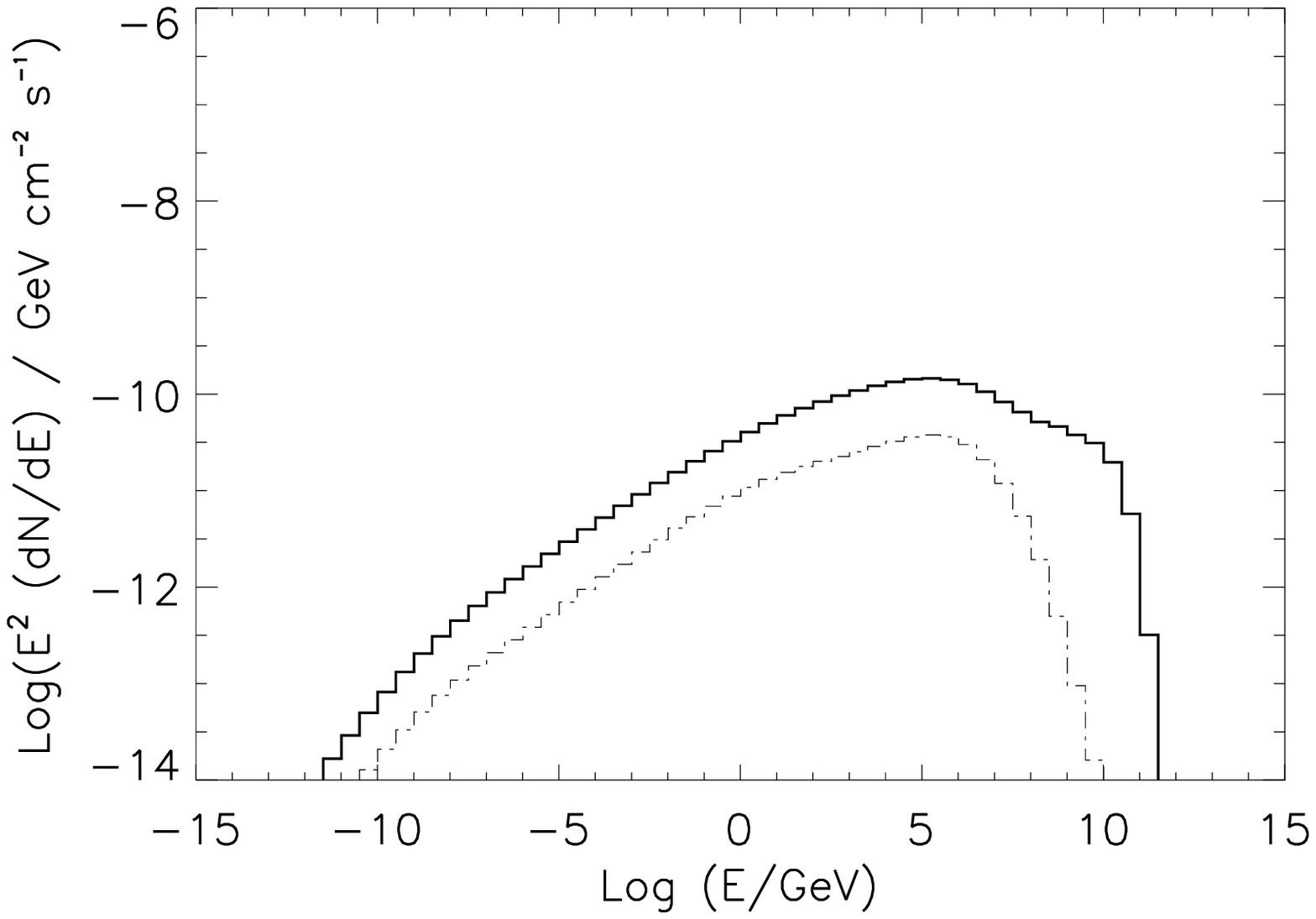,height=2.8in,width=2.9in}
\hfill
\epsfig{file=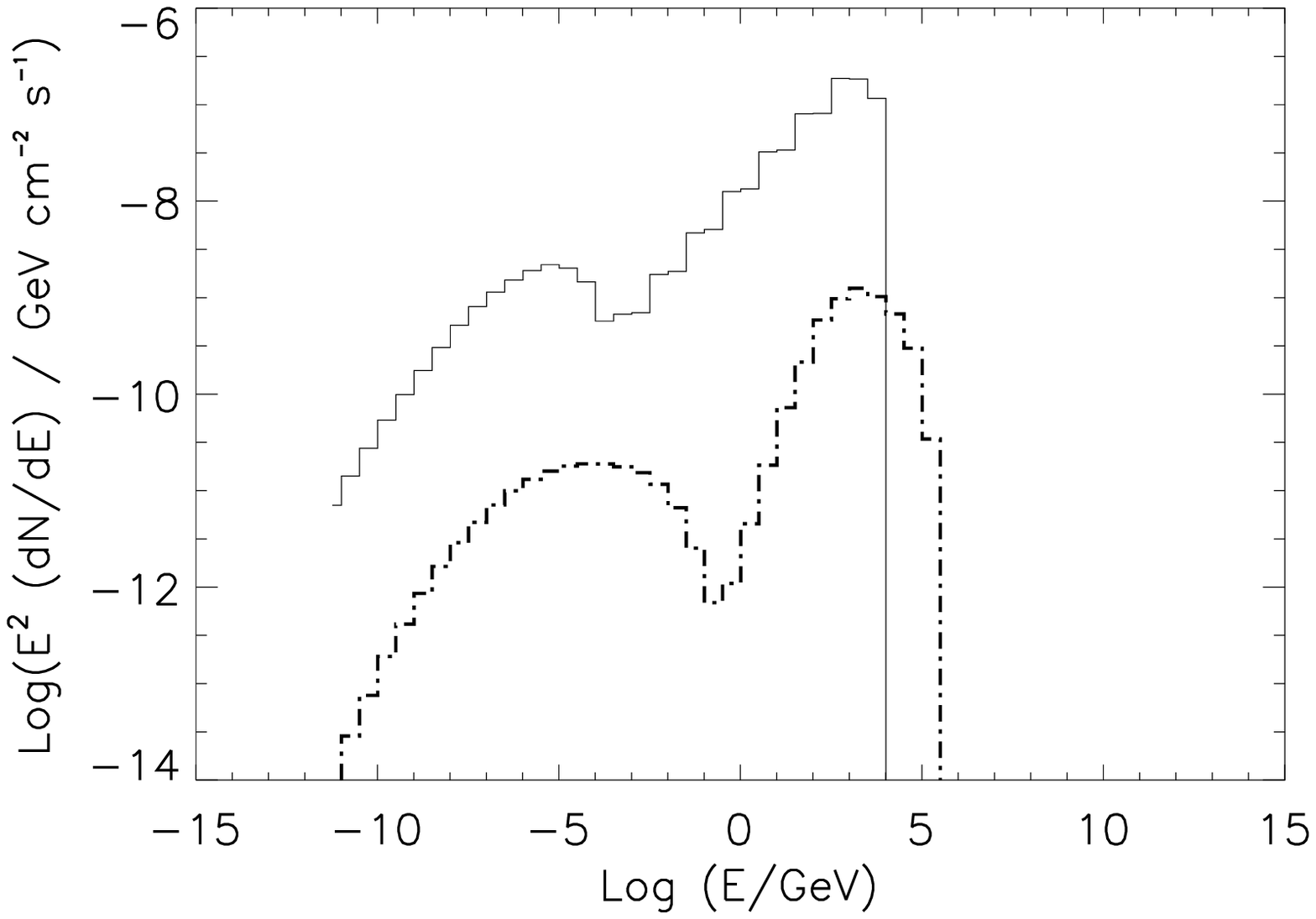,height=2.8in,width=2.9in}
\caption{Left: Average emerging cascade spectra initiated by $\pi^0$-decay (solid line) and $\pi^\pm$-decay
synchrotron photons (dashed-dotted line). Right: Average emerging cascade spectra initiated by
$p$- (solid line) and $\mu^\pm$-synchrotron photons (dashed-dotted line).}
\label{fig3}
\end{figure}

The matrix method (e.g. Protheroe \& Johnson 1996) is used to follow the pair-synchrotron cascade in the ambient synchrotron 
radiation field and magnetic field, 
developing as a result of photon-photon pair production. The cascade can be initiated by 
photons from $\pi^0$-decay (``$\pi^0$-cascade''), electrons from the $\pi^\pm\rightarrow \mu^\pm\rightarrow e^\pm$-decay (``$\pi^\pm$-cascade''), $e^\pm$ from the proton-photon Bethe-Heitler pair production (``Bethe-Heitler-cascade'')
and $p$ and $\mu$-synchrotron photons (``$p$-synchrotron cascade'' and ``$\mu^\pm$-synchrotron cascade'').
In this model, the cascades develop linearly. 

Fig.~3 shows an example of cascade spectra initiated by photons of different origin, and for the
parameter combination given above.
$\pi^0$- and $\pi^\pm$-cascades obviously produce featureless spectra
whereas $p$- and $\mu^\pm$-synchrotron cascades cause the typical double hump shaped SED as 
observed in $\gamma$-ray blazars (see also Rachen, these proceedings).
The contribution from Bethe-Heitler cascades turns out
to be negligible.
Direct $p$- and $\mu^\pm$-synchrotron
radiation is responsible for the high energy peak, whereas the low energy hump may be either
synchrotron radiation from the directly accelerated $e^-$ and/or by pairs produced by the ``low energy hump''.

Adding the four components of the cascade spectrum in Fig.~3 and normalizing
to an ambient, accelerated $p$ density of $n_{tot,p} = 7~\rm{cm}^{-3}$, we obtain the SED shown in Fig.~4
where it is compared with the multifrequency observations of the 16 April 1997 flare of Mkn~501. 

\begin{figure}[h] 
\centerline{\epsfig{file=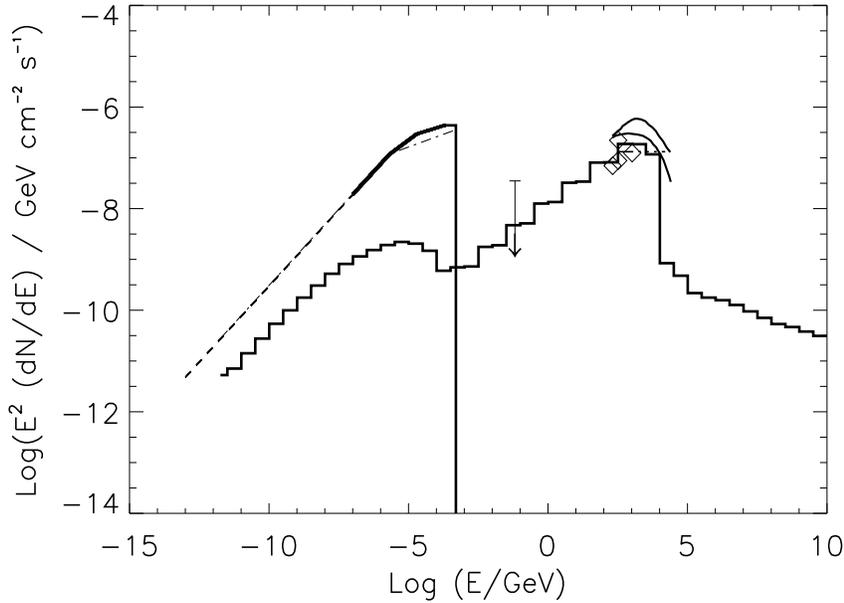,height=3.2in,width=4.5in}}
\caption{
Present model (histogram) in comparison with the data of the 16 April 1997-flare of Mkn~501. Photon
absorption on the cosmic diffuse background radiation field is not included in the model. Straight solid lines:
parametrization of the observed, curved synchrotron spectrum (BeppoSAX \& OSSE) by Bednarek \& Protheroe (1999) 
and observed TeV-emission
corrected for cosmic background absorption (Bednarek \& Protheroe 1999) with peak energy output $\sim 2$~TeV;
the 100~MeV upper limit is from Catanese et al 1997 (observed 9-15 April 1997), diamonds: nearly simultaneous (uncorrected) Whipple data (Catanese et al 1997);
dashed-dotted line: synchrotron target spectrum.
}
\label{fig4}
\end{figure}

\end{document}